\documentstyle[11pt,amssymb]{article}

\def\dalemb#1#2{{\vbox{\hrule height .#2pt
        \hbox{\vrule width.#2pt height#1pt \kern#1pt
                \vrule width.#2pt}
        \hrule height.#2pt}}}
\def\square{\mathord{\dalemb{6.8}{7}\hbox{\hskip1pt}}}

\def\cF{{\cal F}}
\def\cA{{\cal A}}

\def\0{{\sst{(0)}}}
\def\1{{\sst{(1)}}}
\def\2{{\sst{(2)}}}
\def\3{{\sst{(3)}}}
\def\4{{\sst{(4)}}}
\def\5{{\sst{(5)}}}
\def\6{{\sst{(6)}}}
\def\7{{\sst{(7)}}}
\def\8{{\sst{(8)}}}

\def\R{\rlap{\rm I}\mkern3mu{\rm R}}

\def\ep{\epsilon}
\def\td{\tilde}
\def\wtd{\widetilde}

\let\a=\alpha

\def\nn{\nonumber} \def\bd{\begin{document}} \def\ed{\end{document}}
\def\ds{\documentstyle} \let\fr=\frac \let\bl=\bigl \let\br=\bigr
\let\Br=\Bigr \let\Bl=\Bigl 
\let\bm=\bibitem
\let\na=\nabla
\let\pa=\partial \let\ov=\overline 
\newcommand{\be}{\begin{equation}} 
\newcommand{\ee}{\end{equation}} 
\def\ba{\begin{array}}
\def\ea{\end{array}}
\def\ft#1#2{{\textstyle{{\scriptstyle #1}\over {\scriptstyle #2}}}}
\def\fft#1#2{{#1 \over #2}}
\def\del{\partial}
\def\sst#1{{\scriptscriptstyle #1}}
\def\oneone{\rlap 1\mkern4mu{\rm l}}
\def\ie{{\it i.e.\ }}
\def\etc{{\it etc.\ }}
\def\via{{\it via}}
\def\semi{{\ltimes}}
\def\cv{{\cal V}}
\def\str{{\rm str}}
\def\jm{{\rm j}}
\def\im{{\rm i}}
\def\mapright#1{\smash{\mathop{-\!\!\!-\!\!\!-\!\!\!-\!\!\!-\!\!\!
             \longrightarrow}\limits^{#1}}}
\def\maprightt#1#2{\smash{\mathop{-\!\!\!-\!\!\!-\!\!\!-\!\!\!-\!\!\!
             \longrightarrow}\limits^{#1}_{#2}}}

\def\tX{{{\wtd X}}}
\newcommand{\ho}[1]{$\, ^{#1}$}
\newcommand{\hoch}[1]{$\, ^{#1}$}
\newcommand{\bea}{\begin{eqnarray}} 
\newcommand{\eea}{\end{eqnarray}} 
\newcommand{\ra}{\rightarrow}
\newcommand{\lra}{\longrightarrow}
\newcommand{\Lra}{\Leftrightarrow}
\newcommand{\ap}{\alpha^\prime}
\newcommand{\bp}{\tilde \beta^\prime}
\newcommand{\tr}{{\rm tr} }
\newcommand{\Tr}{{\rm Tr} } 
\newcommand{\NP}{Nucl. Phys. }
\newcommand{\tamphys}{\it Center for Theoretical Physics\\
Texas A\&M University, College Station, Texas 77843}
\newcommand{\upenn}{\it Department of Physics and Astronomy\\
University of Pennsylvania, Philadelphia, Pennsylvania 19104}

\newcommand{\auth}{M. Cveti\v{c}\hoch{\dagger1}, 
H. L\"u\hoch{\dagger1} and C.N. Pope\hoch{\ddagger2}
 }

\begin{document}
\begin{flushright}
\hfill{CTP TAMU-42/99}\\
\hfill{UPR-864-T}\\
\hfill{hep-th/9910252}\\
\hfill{October 1999}\\
\end{flushright}

\vspace{15pt}

\begin{center}
{ \large {\bf Four-dimensional $N=4$, $SO(4)$ Gauged Supergravity
from $D=11$}}

\vspace{15pt}
\auth

\vspace{15pt}

{\hoch{\dagger}\upenn}

\vspace{15pt}
{\hoch{\ddagger}\tamphys}

\vspace{40pt}

\underline{ABSTRACT}
\end{center}

We construct the complete and explicit non-linear Kaluza-Klein Ansatz
for deriving the bosonic sector of the standard $N=4$ $SO(4)$ gauged
four-dimensional supergravity from the reduction of $D=11$
supergravity on $S^7$.  This provides a way of interpreting all
bosonic solutions of the four-dimensional gauged theory as exact
solutions in eleven-dimensional supergravity.  We discuss certain
limiting forms of the Kaluza-Klein reduction, and compare them with
related forms in the Freedman-Schwarz $N=4$ $SU(2)\times SU(2)$ gauged
theory.  This leads us to the result that the Freedman-Schwarz model
is in fact a singular limiting case of the standard $SO(4)$ gauged
supergravity.  We show that in this limit, our Ansatz for getting the
$SO(4)$ gauged theory as an $S^7$ reduction from $D=11$ indeed reduces
to an $S^3\times S^3$ reduction from $D=10$, which makes contact with
previous results in the literature.  We also show that there is no
distinction to be made between having equal or unequal values for the
gauge coupling constants $g$ and $\td g$ of the two $SU(2)$
gauge-group factors in the standard $N=4$ $SO(4)$ gauged supergravity,
whilst by contrast the ratio of $g$ to $\td g$ is a non-trivial
parameter of the Freedman-Schwarz model.

{\vfill\leftline{}\vfill
\vskip 5pt
\footnoterule
{\footnotesize \hoch{1} Research supported in part by DOE grant 
DE-FG02-95ER40893 \vskip -12pt} \vskip 14pt
%{\footnotesize \hoch{2} Research supported in part by DOE grant
%DE-AC02-76ER03071 \vskip -12pt} \vskip 14pt
{\footnotesize  \hoch{2} Research supported in part by DOE 
grant DE-FG03-95ER40917.\vskip  -12pt}}

\pagebreak
\setcounter{page}{1}

\section{Introduction}

     In the conjectured AdS/CFT correspondence \cite{malda,gkp,wit}, it
becomes important to establish how the lower-dimensional gauged
supergravities arise through spherical Kaluza-Klein reductions of the
fundamental theories in $D=10$ or $D=11$.  It has long been known in
certain cases that at the level of linearised fluctuations around an
AdS $\times$ Sphere background, the massless excitations in the
Kaluza-Klein spectrum describe the maximal gauged supergravity
multiplet.  The cases where this occurs include $D=11$ supergravity
compactified on $S^7$ \cite{dp} or $S^4$ \cite{PTV}, and type IIB
supergravity compactified on $S^5$ \cite{gunaydin,kim}.

    What is much less clear is whether these results extend nicely
beyond the level of the linearised analysis.  It is obvious that if
one performs expansions of all the fields in terms of complete sets of
harmonics on the sphere, then one will necessarily obtain a
lower-dimensional theory comprising the gauged supergravity coupled to
an infinite tower of massive multiplets.  {\it A priori}, one might
expect that beyond the linearised level, there could be couplings of
the form $H\, L^2$, $H\, L^3$, \etc in the lower-dimensional
Lagrangian, where $H$ represents a heavy field and $L$ a massless one.
Such couplings would prevent one from rigorously setting the heavy
fields to zero, since the massless fields would be acting as sources
for them.  Such a phenomenon does not happen in a toroidal reduction,
since the torus harmonics associated with the massless modes are
constants, while those associated with the massive modes depend on the
torus coordinates.  Thus it is guaranteed in that case that no
non-linear products of zero-mode harmonics can generate non-zero-mode
harmonics.  The truncation to the massless sector is therefore
guaranteed to be consistent in a toroidal reduction.

   On the sphere, the harmonics associated with the massless fields
can depend on the coordinates of the sphere (for example, the Killing
vectors associated with the massless gauge bosons), and so it is far
from obvious that once the non-linear interactions are included, there
will be no couplings linear in heavy fields, of the kind we discussed
above.  Quite the contrary, in fact; it is easy to see that in general
such terms {\it will} be present, and so a generic theory reduced on a
sphere cannot be consistently truncated to the massless sector.
Remarkably, however, it turns out that these consistency problems are
avoided in the case of the sphere reductions of $D=11$ supergravity.
For the $S^7$ reduction, indications of this were seen in \cite{dnpw,wna},
and a complete demonstration of the consistency of the truncation was
given in \cite{deWitnicolai}.  For the $S^4$ reduction, the explicit
reduction Ansatz was recently constructed \cite{nvv}, again showing
that the truncation to the massless sector is consistent.  No
analogous result has been derived for the $S^5$ reduction of type IIB
supergravity, but it is strongly believed to work there too.

    The Kaluza-Klein Ans\"atze for the complete $S^7$ and $S^4$
reductions are rather complicated (the $S^7$ case is especially
complicated, and indeed the reduction scheme obtained in
\cite{deWitnicolai} is somewhat implicit, which is presumably
inevitable since the $N=8$
gauged theory is itself intrinsically rather complicated).  In a
number of recent papers, completely explicit consistent reduction
Ans\"atze have been constructed for various further (consistent)
truncations of the maximal gauged supergravities.  The advantage of
looking at these smaller theories is that the expressions for the
Ans\"atze become much more manageable, and it becomes possible to
present fully explicit results.  These results are completely
sufficient if one is interested in knowing how to embed
lower-dimensional solutions that use only the truncated subset of
fields into the original theory in $D=10$ or $D=11$.  Cases that have
been worked out in this way include truncations to the maximal abelian
subgroups $U(1)^4$, $U(1)^3$ and $U(1)^2$ of the full $SO(8)$, $SO(6)$
and $SO(5)$ gauge groups in $D=4$, 5 and 7 \cite{ten}.  Cases with
surviving non-abelian gauge groups have also been constructed; the
$N=1$ $SU(2)$ gauged supergravity in $D=7$ \cite{d7gauge}, and the
$N=4$ gauged $SU(2)\times U(1)$ supergravity in $D=5$ \cite{d5gauge}.
The former arises from an $S^4$ reduction from $D=11$, while the
latter comes from an $S^5$ reduction from type IIB supergravity.
Another case that has been obtained is $N=2$ $SU(2)$ gauged
supergravity in $D=6$ \cite{d6gauge}.  This is in fact the largest
gauged theory that exists in $D=6$ \cite{romansd6}, even though
ungauged supergravity with $N=4$ exists in $D=6$.  The six-dimensional
theory arises from a local $S^4$ reduction of massive type IIA
supergravity \cite{d6gauge}.\footnote{In all cases, attention has been
focussed on the bosonic sectors of the supergravities, since these are
the fields that participate in $p$-brane solutions.}  An explicit
Ansatz for the embedding
of the symmetric scalar potential of $D=5$ gauged supergravity 
into the metric of type IIB was also obtained, in \cite{fgpw} (see
also \cite{bs}).  This
was extended to the full consistent embedding, giving the Ansatz also
for the 5-form antisymmetric tensor (the only other active field in
this truncation) in \cite{cglp}.  The results were also extended to
the full consistent reduction Ans\"atze for the symmetric scalar potentials
of $D=7$ and $D=4$ gauged supergravities from $D=11$, and for the
analogous scalar potential in $D=6$ gauged supergravity from massive
type IIA, in \cite{cglp}.  

    The consistency of the reduction is of particular importance in
the context of the AdS/CFT correspondence.  One is interested in
considering $p$-brane configurations in the higher dimension that
carry a large charge $N$, in the limit when $N\longrightarrow\infty$.
From the lower-dimensional point of view, this corresponds to
configurations such as charged AdS black holes where the gauge fields
that support the solution take large values.  If, heuristically
speaking, the massless fields denoted by $L$ are taking very large
values then it is crucial that there should be no $H\, L^2$, $H\,
L^3$, \etc, couplings in the theory, in order that the neglect of the
massive fields $H$ can be justified.  

    In this paper, we shall construct another example where an
explicit consistent reduction can be obtained.  We consider the case
of $N=4$ $SO(4)$ gauged supergravity in $D=4$.\footnote{We should
emphasise that here we are, intitially, 
discussing the {\it standard} $N=4$ $SO(4)$ gauged
theory of \cite{dfr}, not the Freedman-Schwarz $N=4$ $SU(2)\times
SU(2)$ gauged theory of \cite{frsw}.  In section 5, however, we show
that the latter is a singular limit of the former.}  
Of course in principle the reduction
Ansatz for this theory should be subsumed in the $N=8$ reduction
described in \cite{deWitnicolai}.  In practice, as we have indicated,
the results in \cite{deWitnicolai} are somewhat implicit, and
furthermore the full results for the reduction of the $D=11$ 4-form
field strength are not presented there.  (The metric reduction Ansatz, on
the other hand, {\it is} given explicitly, and in fact we make use of
the metric reduction given in \cite{deWitnicolai} in obtaining our
results.)  This $N=4$ example is rather more complicated than previous
ones that have been explicitly considered.  In particular, the bosonic field
content includes not only a dilaton but also an axion, and this leads
to a more involved structure in the reduction Ans\"atze.  As usual,
the bulk of the complexity in determining the reduction Ans\"atze
centres around the antisymmetric tensor fields.

    In section 2 we present our results for the consistent
Kaluza-Klein reduction Ansatz, including a discussion of the geometry
of the internal 7-sphere.  In section 3 we present the Lagrangian and
the equations of motion for the four-dimensional $N=4$ $SO(4)$ gauged
supergravity, and in section 4 we discuss how our Kaluza-Klein Ansatz
produces this theory as an exact embedding in $D=11$ supergravity.  In
section 5, we discuss certain singular limits of the reduction, and we
compare them with related limits in the Freedman-Schwarz $N=4$
$SU2)\times SU(2)$ gauged supergravity.  We show that in fact the
Freedman-Schwarz model can be understood as a singular limit of the
standard $N=4$ $SO(4)$ gauged theory, in which the axion is shifted by
an infinite constant.  The $S^7$ internal space degenerates to
$\R\times S^3\times S^3$ in this limit.  We also show that in the
standard $N=4$ $SO(4)$ gauged theory, there is no distinction to be
made between the cases of equal or unequal gauge coupling constants
$g$ and $\td g$ for the two $SU(2)$ factors in the gauge group, since
one can make rescalings that allow the ratio to be adjusted at will.
(This observation was also made in \cite{zw,gazw2}.)
By contrast, no such rescalings are possible in the Freedman-Schwarz
model, and so there the ratio $g/\td g$ is a non-trivial parameter of
the theory.  After concluding remarks, we present details in Appendix
A of the derivation of the metric reduction Ansatz.

\section{The Ansatz}

    In this section, we present our results for the Kaluza-Klein
reduction Ansatz for obtaining $N=4$ $SO(4)$ gauged supergravity in
$D=4$ from an $S^7$ reduction of $D=11$ supergravity.  Some of the
details of how we arrived at this Ansatz are discussed in Appendix A.
For the metric, we find
%%%%%
\be
d\hat s_{11}^2 = \Delta^{\fft23}\, ds_4^2 + 2 g^{-2}\,
\Delta^{\fft23}\, d\xi^2 
+ \ft12 
g^{-2}\, \Delta^{\fft23}\, \Big[\fft{c^2}{c^2\, X^2 + s^2}\, \sum_i (h^i)^2
+ \fft{s^2}{s^2\, \tX^2 + c^2}\, \sum_i (\td h^i)^2\Big]\,,\label{metans}
\ee
%%%%%
where
%%%%%
\bea
&&\tX \equiv  X^{-1}\, q\,,\qquad q^2 \equiv 1 + \chi^2\,
X^4\,,\nn\\
&&\Delta \equiv  \Big[(c^2\, X^2 + s^2)(s^2\, \tX^2 + c^2)
\Big]^{\fft12} \,,\label{defs1}\\
&&c\equiv \cos\xi\,,\qquad s\equiv \sin\xi\,,\nn\\
&& h^i \equiv \sigma_i - g\, A_\1^i\,,\qquad \td h^i \equiv \td\sigma_i
-g\, \wtd A_\1^i\,.\nn
\eea
%%%%%
The three quantities $\sigma_i$ are left-invariant 1-forms on
$S^3=SU(2)$, and the three $\td\sigma_i$ are left-invariant 1-forms
on a second $S^3$.    They satisfy
%%%%%
\be
d\sigma_i = -\ft12 \ep_{ijk}\, \sigma_j\wedge \sigma_k\,,\qquad
d\td\sigma_i = -\ft12 \ep_{ijk}\, \td\sigma_j\wedge \td\sigma_k\,.
\label{leftinv}
\ee
%%%%%
The $SU(2)$ Yang-Mills potentials $A_\1^i$, together with the second
set $\wtd A_\1^i$, together comprise the $SO(4)\sim SU(2)\times SU(2)$
gauge group of the $N=4$ gauged supergravity in $D=4$.  The constant
$g$ is the gauge coupling constant.  The remaining
bosonic fields of the $N=4$ supermultiplet are the dilaton
$\phi$ and the axion $\chi$.  The dilaton parameterises the quantity
$X$ appearing in (\ref{metans}) and (\ref{defs1}), being related to
it by
%%%%%
\be
X= e^{\fft12\phi}\,.
\ee
%%%%%
 
   We find that the Ansatz for $\hat F_\4$ is as follows:
%%%%%
\be
\hat F_\4 = -g\, \sqrt2\, U\, \ep_\4 - 
\fft{4s\, c}{g\,\sqrt2}\, X^{-1}\, {*dX}\wedge
d\xi + \fft{\sqrt2 s\, c}{g}\, \chi\, X^4\, {*d\chi}\wedge d\xi + 
\hat F_\4' + \hat F_\4''\,,\label{fans1}
\ee
%%%%%
where 
%%%%%
\be
U = X^2\, c^2 + \tX^2\, s^2  + 2 \,,
\ee
%%%%%
and $\hat F_\4' = d\hat A_\3'$, with
%%%%%
\be
\hat A_\3' = f\, \ep_\3 + \td f\, \td\ep_3\,,\label{a3p}
\ee
%%%%%
where $\ep_\3 = \ft16 \ep_{ijk} \, h^i\wedge h^j\wedge h^k$ and
$\td\ep_\3 = \ft16 \ep_{ijk} \, \td h^i\wedge \td h^j\wedge \td h^k$.
The functions $f$ and $\td f$ are given by
%%%%%
\bea
f &=& \fft1{2\sqrt2}\, g^{-3}\, c^4 \chi\, X^2\, (c^2\, X^2+s^2)^{-1}\,,\nn\\
\td f &=& -\fft1{2\sqrt2}\, 
g^{-3}\, s^4\, \chi\, X^2\,  (s^2\, \tX^2 + c^2)^{-1}\,.
\label{ftf}
\eea
%%%%%
The field strength contribution $\hat F_\4'$ is therefore given by
%%%%%
\bea
\hat F_\4' &=& \fft{\del f}{\del\chi}\, d\chi\wedge \ep_\3 + \fft{\del
f}{\del X}\, dX\wedge \ep_\3 + \fft{\del f}{\del \xi}\, d\xi \wedge
\ep_\3 \nn\\
&&+  \fft{\del \td f}{\del\chi}\, d\chi\wedge \td\ep_\3 + \fft{\del
\td f}{\del X}\, dX\wedge \td\ep_\3 + \fft{\del \td f}{\del \xi}\, d\xi \wedge
\td\ep_\3\nn\\
&&-\ft12 f\, g\, \ep_{ijk}\, h^i\wedge h^j\wedge F_\2^k -
 \ft12 \td f\, g\, \ep_{ijk}\, \td h^i\wedge \td h^j\wedge \wtd
 F_\2^k\,.\label{fans2}
\eea
%%%%%

    The terms in $\hat F_\4''$ comprise those involving the
$SU(2)\times SU(2)$ Yang-Mills field strengths $F_\2^i$ and $\wtd
F_\2^i$.  These are given by
%%%%%
\bea
\sqrt2\, \hat F_\4''&=& g^{-2}\, s\, c\, X^{-2}\, d\xi\wedge h^i\wedge
{*F_\2^i} +\ft14 g^{-2}\, c^2\, X^{-2}\, \ep_{ijk}\, h^i\wedge
h^j\wedge {*F_\2^k}\nn\\
&&-g^{-2}\, s\, c\, \tX^{-2}\, d\xi\wedge \td h^i\wedge
{*\wtd F_\2^i} +\ft14 g^{-2}\, s^2\, \tX^{-2}\, \ep_{ijk}\, \td h^i\wedge
\td h^j\wedge {*\wtd F_\2^k}\,,\nn\\
&&+g^{-2}\, s\, c\, \chi\, d\xi\wedge h^i\wedge F_\2^i +
\ft14 g^{-2}\, c^2 \chi\,\ep_{ijk}\, h^i\wedge h^j\wedge F_\2^k \,,
\label{fpp}\\
&&+g^{-2}\, s\, c\, \chi\, X^2\,\tX^{-2}\, d\xi\wedge \td h^i
\wedge \wtd F_\2^i -
\ft14 g^{-2}\, s^2\, \chi\, X^2\,\tX^{-2}\, \ep_{ijk}\,
\td h^i\wedge \td h^j\wedge \wtd F_\2^k\,.\nn
\eea
%%%%%

   For the purposes of verifying the consistency of the Ansatz, it is
useful to record that the eleven-dimensional Hodge dual of $\hat F_\4$
is given by
%%%%%
\bea
{\hat *\hat F_4} &=& \ft14 g^{-6}\, s^3\, c^3\, \Delta^{-2}\, U\, d\xi\wedge
\ep_\3 \wedge \td\ep_\3 - \ft14 g^{-6}\, s^4\, c^4\,\Delta^{-2}\,
X^{-1}\, dX\wedge \ep_\3\wedge \td\ep_\3 \nn\\
&&+\ft18 g^{-6}\, s^4\, c^4\,
\Delta^{-2}\, X^4\, \chi\, d\chi\wedge\ep_\3 \wedge \td\ep_\3
+ {\hat *\hat F_\4'} + {\hat *\hat F_\4''}\,,\label{hodge1}
\eea
%%%%%
where the term ${\hat *\hat F_\4'}$ is given by
%%%%%
\bea
{\hat *\hat F_\4'} &=& -\sqrt2\, g^{-1}\, s^3\, c^{-3}\, 
\fft{\del f}{\del \chi}\, \Delta^{-2}\, \Omega^3\, {*d\chi}\wedge d\xi \wedge
\td\ep_\3\nn\\
&& -\sqrt2\, g^{-1}\, s^3\, c^{-3}\, 
\fft{\del f}{\del X}\, \Delta^{-2}\, \Omega^3\, {*dX}\wedge d\xi \wedge
\td\ep_\3\nn\\
&& + \sqrt2\, g^{-1}\, c^3\, s^{-3}\, 
\fft{\del \td f}{\del \chi}\, \Delta^{-2}\, \wtd\Omega^3\, 
{*d\chi}\wedge d\xi \wedge \ep_\3 \nn\\
&& + \sqrt2\, g^{-1}\, c^3\, s^{-3}\, 
\fft{\del \td f}{\del X}\, \Delta^{-2}\, \wtd\Omega^3\,  
{*dX}\wedge d\xi \wedge \ep_\3\nn\\
&&+\ft1{\sqrt2}\,  
g\, s^3\, c^{-3}\, \fft{\del f}{\del\xi}\, \Delta^{-2}\, \Omega^3\,
 \ep_\4\wedge \td \ep_\3 
-\ft1{\sqrt2} \, g\, c^3\, s^{-3}\, \fft{\del \td f}{\del\xi}\, \Delta^{-2}\, 
\wtd\Omega^3\, \ep_\4\wedge \ep_\3\\
&&-\ft1{\sqrt2}\, f\, g^{-2}\, s^3\, c^{-1}\, \Omega\,
\wtd\Omega^{-1}\, d\xi \wedge h^i \wedge {*F_\2^i}\wedge \td\ep_\3\nn\\
&& + \ft1{\sqrt2}\, \td f\, g^{-2}\, c^3\, s^{-1}\, \wtd \Omega\,
\Omega^{-1}\, d\xi \wedge \td h^i \wedge {*\wtd F_\2^i}\wedge \ep_\3\,,\nn
\eea
%%%%%
and we have defined
%%%%%
\be
\Omega \equiv c^2\, X^2 + s^2\,,\qquad 
\wtd\Omega \equiv s^2\, \tX^2 + c^2\,.
\ee
%%%%%
The final term in (\ref{hodge1}) is given by
%%%%%
\bea
{\hat *\hat F_\4''} &=& -\ft1{16} g^{-5}\, s^4\, c^2\, \wtd\Omega^{-1}\,
X^{-2}\, \ep_{ijk}\, h^j\wedge h^k\wedge F_\2^k\wedge \td \ep_\3\nn\\
&& -\ft14g^{-5}\, s^3\, c\, \Omega\, \wtd\Omega^{-1}\, X^{-2}\, d\xi\wedge
h^i\wedge F_\2^i \wedge \td\ep_3\nn\\
&&
 -\ft1{16} g^{-5}\, s^2\, c^4\, \Omega^{-1}\,
\tX^{-2}\, \ep_{ijk}\, \td h^j\wedge \td h^k\wedge \wtd F_\2^k\wedge \ep_\3
\nn\\
&& +\ft14 g^{-5}\, s\, c^3\, \wtd\Omega\, \Omega^{-1}\, \tX^{-2}\, d\xi\wedge
\td h^i\wedge \wtd F_\2^i \wedge \ep_3\nn\\
&& +\ft1{16} g^{-5}\, s^4\, c^2\, \wtd\Omega^{-1}\,
\chi\, \ep_{ijk}\, h^j\wedge h^k\wedge {*F_\2^k}\wedge \td \ep_\3
\\
&& +\ft14 g^{-5}\, s^3\, c\, \Omega\, \wtd\Omega^{-1}\, \chi\, d\xi\wedge
h^i\wedge {*F_\2^i} \wedge \td\ep_3\nn\\
&& -\ft1{16} g^{-5}\, s^2\, c^4\, \Omega^{-1}\,\chi\, X^2
\tX^{-2}\, \ep_{ijk}\, \td h^j\wedge \td h^k\wedge 
{* \wtd F_\2^k}\wedge \ep_\3\nn\\
&&+\ft14 g^{-5}\, s\, c^3\, \wtd\Omega\, \Omega^{-1}\, \chi\, X^2\, 
\tX^{-2}\, d\xi\wedge
\td h^i\wedge {*\wtd F_\2^i} \wedge \ep_3\nn
\eea
%%%%%
  
    A number of remarks about the reduction Ansatz are in order.
First, we note that there is a residual $Z_2$ subgroup of the original
global $SL(2,\R)$ symmetry of the ungauged theory, under which the
various quantities are mapped to their primed images, given by
%%%%%
\bea
&&X'=\tX\,,\qquad \chi'\, X'^2 = -\chi\, X^2\,,
\qquad {A'}^i_\1 = \wtd A^i_\1\,,\qquad
\td {A'}^i_\1 = A^i_\1\,,\nn\\
&&c'=s\,,\qquad s'=-c\,,\qquad {{h_i}}'=\td h_i\,,\qquad {{\td h_i}}' =
h_i\,,\label{z2sym}\\
&&\ep_\3' = \td \ep_\3\,,\qquad \td\ep{\,}'_\3 = \ep_\3\,.\nn
\eea
%%%%%
In particular, we have $\Delta'=\Delta$, $q'=q$ and $U'=U$.  In fact, the
entire Ansatz for the metric and 4-form is invariant under the $Z_2$.
It corresponds to an interchange of the two 3-spheres in our
description of $S^7$ as a foliation of $S^3\times S^3$.
Correspondingly, in the four-dimensional theory itself, the $Z_2$
symmetry involves an interchange of the two $SU(2)$ gauge fields.

    The geometry of the internal 7-sphere can be understood as
follows.  If we look at the metric Ansatz (\ref{metans}) in the
``unexcited'' state where the gauge fields, axion and dilaton vanish 
(and so $X=\tX=1$), we see that up to a constant factor of $\ft12
g^{-2}$ the internal 7-dimensional metric becomes
%%%%%
\be
d\Omega_7^2 =  d\xi^2 + \cos^2\xi\, d\Omega_3^2 + \sin^2\xi\,
d\wtd\Omega_3^2\,,
\ee
%%%%%
where $d\Omega_3^2=\ft14 \sum_i \sigma_i^2$ and $d\wtd\Omega_3^2=\ft14
\sum_i \td\sigma_i^2$ are metrics on the two unit 3-spheres.  In fact
$d\Omega_7^2$ is a metric on the unit 7-sphere, with the ``latitude''
coordinate $\xi$ running between the limits $0\le\xi\le \ft12\pi$, at
which one or other of the two 3-spheres shrinks to zero radius.  This
geometrical description of the 7-sphere is analogous to the
description of a $S^3$ as a foliation of Clifford tori $S^1\times
S^1$, in which one has $d\Omega_3^2 = d\xi^2 + \cos^2\xi\, d\phi_1^2 +
\sin^2\xi\, d\phi_2^2$.

\section{$N=4$ $SO(4)$ gauged four-dimensional supergravity}

    The bosonic Lagrangian is given by
%%%%%
\bea
{\cal L}_4 &=& R\, {*\oneone} - \ft12 {*d\phi}\wedge d\phi - \ft12
e^{2\phi}\, {*d\chi}\wedge d\chi - V\, {*\oneone} \nn\\
&&- \ft12 e^{-\phi}\, {*F_\2^i}\wedge F_\2^i  -\ft12 \fft{e^\phi}{1+
\chi^2\, e^{2\phi}}\, {* \wtd F_\2^i}\wedge \wtd F_\2^i\,,\label{d4lag}\\
&&- \ft12\chi\, F_\2^i\wedge F_\2^i + 
\ft12 \fft{\chi\, e^{2\phi}}{1+\chi^2\, e^{2\phi}}\, \wtd F_\2^i \wedge
\wtd F_\2^i\,,\nn
\eea
%%%%%
where the potential $V$ is
%%%%%
\be
V = -2g^2\, (4+ 2 \cosh\phi + \chi^2\, e^\phi)\,,\label{pot0}
\ee
%%%%%
and
%%%%%
\be
F_\2^i = dA_\1^i + \ft12 g\, \ep_{ijk}\, A_\1^j\wedge A_\1^k\,,\qquad
\wtd F_\2^i = d\wtd A_\1^i + \ft12 g\, \ep_{ijk}\, \wtd A_\1^j
\wedge \wtd A_\1^k\,.\label{fde0}
\ee
%%%%%
(We have chosen to set the two gauge couplings for the two $SU(2)$
gauge groups equal here.  There is no loss of generality involved in
doing this; one can always restore the two coupling constants by
shifting $\phi$ by a constant, accompanied by appropriate rescalings
of $\chi$ and the gauge potentials.  We shall return to this point 
in section 5.)

   The dilaton $\phi$ is related to the quantity $X$ of the previous
   section by
%%%%%
\be
X= e^{\fft12 \phi}\,.
\ee
%%%%%

   The equations for motion for $X$ and $\chi$ that follow from
(\ref{d4lag}) are:
%%%%%
\bea
d(X^{-1}\, {*dX}) &=& -\ft12 X^4\, {*d\chi}\wedge d\chi + 
 g^2\, (X^2 -X^{-2} + \chi^2\, X^2)\, \ep_\4 
+\ft14 X^{-2}\, {*F_\2^i}\wedge F_\2^i\nn\\
&& -\ft14 (1-\chi^2\, X^4)\,
X^2\, q^{-4}\, {*\wtd F_\2^i}\wedge \wtd F_\2^i + 
\ft12 \chi\, \tX^{-4}\, \wtd F_\2^i\wedge \wtd F_\2^i\,,\\
d(X^4\, {*d\chi}) &=& 4 g^2\, \chi\, X^2\, \ep_\4 + 
\chi\, X^6\, q^{-4}\, {*\wtd F_\2^i}\wedge \wtd F_\2^i\nn\\
&&-\ft12 F_\2^i\wedge F_\2^i + \ft12 (1-\chi^2\, X^4)\, \tX^{-4}\,
\wtd F_\2^i\wedge \wtd F_\2^i\,,
\eea
%%%%%
where we are using the functions $q$ and $\tX$ defined in the previous section.

   The Yang-Mills equations of motion are
%%%%%
\bea
D(X^{-2}\, {*F_\2^i}) &=& -d\chi\wedge F_\2^i \,,\nn\\
\wtd D(\tX^{-2}\, {* \wtd F_\2^i}) &=& d(\chi\, X^2 \, \tX^{-2})\wedge \wtd
F_\2^i\,.
\eea
%%%%%
where $D$ and $\wtd D$ are the Yang-Mills-covariant exterior
derivatives for the two $SU(2)$ gauge groups:
%%%%%
\bea
D\, F_\2^i &\equiv& dF_\2^i + g\, \ep_{ijk}\, A_\1^j\wedge F_\2^k =0\,,\nn\\
\wtd D\, \wtd F_\2^i &\equiv& d\wtd F_\2^i + g\, \ep_{ijk}\, 
\wtd A_\1^j\wedge \wtd F_\2^k =0\,,\nn
\eea
%%%%%
{\it etc}.   Finally, the Einstein equation is
%%%%%
\bea
R_{\mu\nu} &=& \ft12 \del_\mu\phi\, \del_\nu\phi + \ft12 e^{2\phi}\, 
\del_\mu\chi\, \del_\nu\chi + \ft12 X^{-2}\, (F^i_{\mu\rho}\, F_{\nu}^{i\,
\rho} -\ft14 (F^i_\2)^2 \, g_{\mu\nu})\nn\\
&& 
+ \ft12 \tX^{-2}\, (\wtd F^i_{\mu\rho}\, \wtd F_{\nu}^{i\,
\rho} -\ft14 (\wtd F^i_\2)^2 \, g_{\mu\nu})\,.
\eea
%%%%% 

     Note that the $Z_2$ symmetry (\ref{z2sym}) can be seen in the Lagrangian 
(\ref{d4lag}).   It can be made manifest by making use of the $\tX$ 
variable, to rewrite (\ref{d4lag}) as  
\bea
{\cal L}_4 &=& R\, {*\oneone} - \ft12 {*d\phi}\wedge d\phi - \ft12
e^{2\phi}\, {*d\chi}\wedge d\chi - V\, {*\oneone} \nn\\
&&- \ft12 X^{-2}\, {*F_\2^i}\wedge F_\2^i  -\ft12 \tX^{-2}\, 
{* \wtd F_\2^i}\wedge \wtd F_\2^i\,,\label{d4lag2}\\
&&- \ft12\chi\, F_\2^i\wedge F_\2^i +
\ft12 \chi\, X^2\, \tX^{-2} \wtd F_\2^i \wedge
\wtd F_\2^i\,,\nn
\eea
%%%%%
where the potential $V$ can be written as 
%%%%%
\be
V = -2 g^2\, (4+ X^2+ \tX^2)\,,\label{pot2}
\ee

\section{Reduction from $D=11$ to $D=4$}

    In section 2 we presented our results for the Ans\"atze for the
metric tensor and 4-form field strength of eleven-dimensional
supergravity, which, when substituted into the eleven-dimensional
equations of motion, give rise to the equations of motion for the
four dimensional $N=4$ $SO(4)$ gauged supergravity.  We arrived at the
Ansatz for the metric by a combination of generalisation from the
previously known abelian case in \cite{ten}, and the general formula
presented in \cite{deWitnicolai}, as described in Appendix A.  Our
procedure for determining the 4-form field strength Ansatz consisted
of a combination of generalisation from the abelian case in
\cite{ten}, together with a trial and error process of introducing
additional terms as necessary until consistency was achieved.  Thus
our criterion for determining the Ansatz was to verify explicitly that
substituting it into the eleven-dimensional equations of motion gave
a consistent reduction to the four-dimensional equations of
motion. (In particular, the most non-trivial part is finding an Ansatz
that is {\it consistent}, in the sense that all the dependence on the
$\xi$ coordinate of the 7-sphere cancels out in all the equations.)

    We shall not present all the details of the substitution of the
Ansatz here, because the procedure is an involved one, and in fact
parts of it were most conveniently checked by computer.  However, it
is useful to summarise the structure of the calculation.  The
$D=11$ equations of motion, and the Bianchi identity for $\hat F_\4$, are
given by
%%%%%
\bea
\hat R_{MN} &=& \ft1{12}(\hat F^2_{MN} - 
\ft1{12} \hat F^2_\4\, \hat g_{MN})\,,\nn\\
d{\hat *\hat F_\4}&=& - \ft12 \hat F_\4\wedge \hat F_\4\,,\label{d11eom}\\
d\hat F_\4&=&0\,.\nn
\eea
%%%%%
Considering first the Bianchi identity, it is evident from
(\ref{fans1}), (\ref{fans2}) and (\ref{fpp}) that since the
four-dimensional Hodge duals ${*dX}$, ${*d\chi}$, ${*F_\2^i}$ and
${*\wtd F_\2^i}$ appear in the Ansatz for $F_\4$, it must be that
$d\hat F_\4=0$ will not be satisfied as an {\it identity}, but rather
it will imply certain of the four-dimensional equations of motion.
Specifically, $d\hat F_\4=0$ implies the $D=4$ Yang-Mills equations,
and a particular combination of the equations of motion for the
dilaton and the axion.

    The $D=11$ field equation for $\hat F_\4$ gives rise separately to the
four-dimensional equations of motion for the dilaton, the axion, and the 
Yang-Mills fields.  Finally, the various components of the
eleven-dimensional Einstein equation give rise again to the
four-dimensional dilaton, axion and Yang-Mills equations, and also the
four-dimensional Einstein equation.  

   Two comments are in order.  Firstly, we remark that, as always in
these examples of non-trivial consistent sphere reductions, the
consistency is achieved only because of remarkable ``conspiracies''
between the contributions from the metric and the antisymmetric tensor
in the higher-dimensional theory.  Thus in this case it is only
because of the precise field content, and the structure, of the
eleven-dimensional theory that it is possible to obtain a consistent
reduction Ansatz in which all the dependence on the coordinates of the
internal 7-sphere cancels out when the Ansatz is substituted into the
eleven-dimensional equations of motion.

    The second comment is that, as in most of the other cases of
consistent sphere reductions, we see here also that the Ansatz for the 
antisymmetric tensor must be made on the {\it field strength} $\hat
F_\4$, rather than on the fundamental potential $\hat A_\3$.  This is
evident from the fact that the four-dimensional Hodge duals of $dX$,
$d\chi$, $F_\2^i$ and $\wtd F_\2^i$ appear in the Ansatz
(\ref{fans1}), (\ref{fpp}) (as well as the undualised fields).  As we
remarked above, this means that the Bianchi identity for $\hat F_\4$
is not satisfied {\it identically}, but rather as a consequence of the
four-dimensional equations of motion.
Consequently, there is no way to write an explicit Ansatz for the
potential $\hat A_\3$, since if we could, $d\hat F_\4=0$ would be a
true identity.  The upshot from this is that the Kaluza-Klein sphere
reduction must necessarily be discussed at the level of the
higher-dimensional equations of motion; it is not possible to
describe the reduction at the level of substituting an Ansatz into the
higher-dimensional Lagrangian.  

\section{Gauge-coupling limits}

    In this section, we address two main topics.  Firstly, we show how
in the standard $N=4$ $SO(4)$ gauged supergravity, the case with
independent $SU(2)$ coupling constants $g$ and $\td g$ can be derived
from the the case where $g=\td g$, just by field redefinitions.  Thus
there is really no greater generality when the coupling constants are
unequal, in the standard $SO(4)$ gauged theory.  It is, however,
useful to introduce the artificial extra parameter for the purpose of
discussing singular limits.  In the second part of this section, we
first observe that if one or other of the $SU(2)$ coupling constants
is set to zero in the standard $SO(4)$ gauged supergravity, the theory
becomes equivalent to the the similar limit of the Freedman-Schwarz
$SU(2)\times SU(2)$ gauged theory.  Then, we show a more surprising
result, which is that the full Freedman-Schwarz theory with $g$ and
$\td g$ both non-zero can in fact be derived as a singular limit of
the standard $SO(4)$ gauged theory.  This is a limit where the axion
$\chi$ is shifted by an infinite constant, accompanied by appropriate
constant rescalings of certain other fields and coupling constants.
We show how in this limit, the previous $S^7$ reduction Ansatz reduces
to an $\R\times S^3\times S^3$ reduction, which can be interpreted as
an $S^3\times S^3$ reduction from $D=10$.  This makes contact with
previous results \cite{vc,cham} for obtaining the Freedman-Schwarz model by
Kaluza-Klein reduction.

\subsection{$N=4$ $SO(4)$ gauged supergravity with $g\ne \td g$ from
$g=\td g$} 

    To begin, let us show how we can restore the two independent gauge
coupling constants $g$ and $\td g$ in the $N=4$ $SO(4)$ gauged theory,
one for each $SU(2)$ factor in the gauge group.  To do this, we take
the Lagrangian (\ref{d4lag}), and make the following field and
coupling constant redefinitions:
%%%%%
\bea
&&\phi=\phi' + k\,,\qquad \chi= \chi'\, e^{-k}\,,\qquad 
A_\1^i = {A'}^i_\1 \, e^{\fft12 k }\,,\qquad
\wtd A_\1^i = \td {A'}_\1^i\, e^{-\fft12 k}\,,\nn\\
&&g' = g\, e^{\fft12 k}\,,\qquad 
\td g' = g\, e^{-\fft12 k}\,,\label{scaling}
\eea
%%%%%
where $k$ is a constant.  Now dropping the primes, we find that the
Lagrangian takes the identical form (\ref{d4lag}), where now
(\ref{pot0}) and (\ref{fde0}) have become
%%%%%
\be
V = -8 g\, \td g\, - 2g^2\, e^\phi - 2 \td g^2\, e^{-\phi} -
2 \td g^2\, \chi^2\, e^\phi\,,\label{pot00}
\ee
%%%%%
and
%%%%%
\be
F_\2^i = dA_\1^i + \ft12 g\, \ep_{ijk}\, A_\1^j\wedge A_\1^k\,,\qquad
\wtd F_\2^i = d\wtd A_\1^i + \ft12 \td g\, \ep_{ijk}\, \wtd A_\1^j
\wedge \wtd A_\1^k\,.\label{fde00}
\ee
%%%%%
The potential (\ref{pot00}) can also be rewritten in various equivalent
ways: 
%%%%%
\bea
V&=& - 8 g\, \td g  - 2 g^2\, X^2 - 2\td g^2\, \tX^2\,,\nn\\
&=& -8 g\, \td g - 2(g^2 + \td g^2)\, \cosh\lambda - 2(g^2 -\td g^2)\, 
\cos\sigma\, \sinh\lambda\,,\label{various}\\
&=& -\fft1{1-|W|^2}\, \Big( g_+^2\, (3-|W|^2) - g_-^2\, (1-3 |W|^2) -
4 g_+\, g_-\, A\Big)\,.\nn
\eea
%%%%
In the second line, we are using the parametrisation of the scalar
fields given by (\ref{siglam}) in Appendix A.  
In the final line, we have written
the potential in terms of the complex field $W=-A+\im\, B$ used in
\cite{gazw}, which is related to our $\sigma$ and $\lambda$ by  
$W= e^{\im\,\sigma}\, \tanh\ft12\lambda$, and coupling constants $g_\pm= g\pm
\td g$.  

   Thus after the rescaling (\ref{scaling}), the standard $N=4$
$SO(4)$ gauged supergravity with $g=\td g$ is mapped into the
formulation with two independent gauge coupling constants that was
presented in \cite{gazw}.  (This was also observed in
\cite{gazw2,zw}.)  As we have seen, it is in fact identical,
modulo field redefinitions, to the original theory obtained in
\cite{dfr} where the two gauge coupling constants were equal.  (It is
easy to see from (\ref{various}) 
how this equivalence can pass unnoticed if one uses the
$(\sigma,\lambda)$ or $W=-A+\im\, B$ parametrisation for the scalar
fields.)  It is, of course, trivial to substitute the rescalings into
the metric and 4-form Ans\"atze given in section 2, to obtain
reduction Ans\"atze where the two gauge coupling constants are
different.

\subsection{Freedman-Schwarz as a limit of $N=4$
$SO(4)$ gauged supergravity}

    Having obtained the $N=4$ $SO(4)$ gauged supergravity with
independent $SU(2)$ coupling constants $g$ and $\td g$, the
possibility of taking interesting singular limits arises.  First, we
may consider the situation where we set $\td g=0$, whereupon the
potential $V$ becomes
%%%%%
\be
V = -2 g^2\, e^\phi\,.
\ee
%%%%%
Of course the $SU(2)$ gauge fields $\wtd A_\1^i$ just become abelian
$U(1)^3$ in this limit.  The theory in this limit is equivalent to the
limit of the Freedman-Schwarz model in which one of its $SU(2)$ gauge
coupling constants is also set to zero.  (This observation was also
made in \cite{gazw}.)  The equivalence can be made explicit by
dualising the fields $\wtd A_\1^i$ (which can now be done because they
are abelian).  This gives precisely the $\td g=0$ limit of the
Freedman-Schwarz model (see equation (\ref{fslag}) below).  Note that
instead taking the limit where $g=0$ rather than $\td g=0$ is
equivalent, after a field redefinition.

   One might now wonder if it could be possible to obtain the complete
Freedman-Schwarz model, with both gauge coupling constants non-zero,
as a suitable limit of the standard $N=4$ $SO(4)$ gauged supergravity
given by (\ref{d4lag}).  As we shall now show, this is indeed the
case.  Let us first present the bosonic Lagrangian for the
Freedman-Schwarz theory \cite{frsw}: 
%%%%%
\bea 
{\cal L}_4^{\sst{FS}} &=&
R\, {*\oneone} - \ft12 {*d\phi}\wedge d\phi - \ft12 e^{2\phi}\,
{*d\chi}\wedge d\chi - V\, {*\oneone} \label{fslag}\\
&&- \ft12 e^{-\phi}\, {*F_\2^i}\wedge F_\2^i  -\ft12 e^{-\phi}
\, {* \wtd F_\2^i}\wedge \wtd F_\2^i
- \ft12\chi\, F_\2^i\wedge F_\2^i - 
\ft12 \chi\, \wtd F_\2^i \wedge \wtd F_\2^i\,,\nn
\eea
%%%%%
with 
%%%%%
\be
V= -2(g^2 + \td g^2)\, e^\phi\,,\label{fspot}
\ee
%%%%%
and gauge field strengths given by (\ref{fde00}).  

    A natural attempt to obtain this as a limit from (\ref{d4lag})
is to redefine the fields and coupling constants in
(\ref{d4lag}), (\ref{pot00}) and (\ref{fde00}) according to
%%%%%
\be
\chi= \chi' + b\,,\qquad \wtd A_\1^i = b\, \td {A'}_\1^i\,,\qquad 
\td g =\td g'\, b^{-1}\,,\label{btrans}
\ee
%%%%%
(with all other fields and constants left unscaled), where $b$ is a
constant.  Indeed, upon sending $b$ to infinity and dropping the
primes, we find that (\ref{d4lag}) becomes precisely (\ref{fslag})
with the potential $V$ given by (\ref{fspot}), and the field strengths
given by (\ref{fde00}).\footnote{Two ostensibly divergent terms of the
form $b\, F_\2^i\wedge F_\2^i$ and $b\, \wtd F_\2^i\wedge \wtd F_\2^i$
are actually total derivatives, which can be dropped.}  (The effect of
reversing the sign of the dilaton coupling from $e^\phi$ to
$e^{-\phi}$ in the kinetic term for $\wtd F_\2^i$, which is normally
accomplished by dualisation, is instead achieved here by this singular
limiting process.)

     Normally, one would say that the standard $N=4$ $SO(4)$ gauged
supergravity and the Freedman-Schwarz $N=4$ $SU(2)\times SU(2)$
gauged supergravity are intrinsically inequivalent.  This can be understood
from the fact that they correspond to gauging two different
formulations of the same $N=4$ ungauged theory, which are related by
a dualisation involving the gauge fields \cite{cfs}.  The processes of
gauging and dualisation do not commute (since one cannot dualise
non-abelian Yang-Mills fields), and so the gauged theories can
no longer be equivalent.  Thus normally, one would say that to
``relate'' the standard $SO(4)$ gauged theory to the Freedman-Schwarz
theory, it would be necessary to ungauge one theory, dualise, and then
regauge it.  

    An intriguing outcome of our work is that there is another way of
achieving the same effect, by taking a singular limit.  Since the
limit {\it is} singular, one should perhaps still view the two
gauged theories as being in some sense inequivalent.  However, one theory
seems to be ``more inequivalent'' than the other, since we can derive
Freedman-Schwarz as a singular limit of the standard $SO(4)$ gauged
theory, but the arrow cannot be reversed.\footnote{Note that one can
also apply a similar singular limiting procedure in other examples,
including ordinary ungauged supergravities.}

    One can also verify that the supersymmetry transformation laws of
the standard $SO(4)$ gauged theory do indeed produce those of the
Freedman-Schwarz gauged theory when the $b\longrightarrow\infty$ limit
of (\ref{btrans}) is taken.  We shall not present all the details
here, but just the ``extra'' terms in the spin-$\ft32$ and
spin-$\ft12$ transformation laws, which appear only in the gauged
theories.  From \cite{gazw}, and using the notation of that paper,
these extra terms in the $SO(4)$ gauged theory are
%%%%%
\bea
\delta'\, \bar\psi^i_\mu &=& \ft{\im}{2}\, \bar\ep^i\, \gamma_\mu\, 
\fft{[g_+ + g_-\,  (-A+\im\, \gamma_5\, B)]}
{ (1-|W|^2)^{\fft12}}\,,\nn\\
\delta'\, \bar\chi^i &=& \ft1{\sqrt2}\, \bar\ep^i\, 
\fft{[g_+\, (A-\im\, \gamma_5\, B) - g_-]}{ (1-|W|^2)^{\fft12}}\,,
\eea
%%%%%
where $A$, $B$ and $g_\pm$ were defined in section 5.1.  Making the
replacements (\ref{btrans}), and then sending $b$ to infinity, we find
that there are exact cancellations of terms linear in $b$, which would
otherwise have been divergent, leaving an overall finite result, namely
%%%%%
\bea
\delta'\, \bar\psi^i_\mu &=& \ft{\im}{2}\, e^{\fft12\phi}\,
\bar\ep^i\, (\td g - \im \, g\, \gamma_5)\, \gamma_\mu\,,\nn\\
\delta'\, \bar\chi^i &=& \ft1{\sqrt2}\, e^{\fft12\phi}\, \bar\ep^i\, 
 (\td g - \im \, g\, \gamma_5)\,.
\eea
%%%%%
These are precisely the correct forms of the corresponding ``extra''
terms in the transformation rules of the Freedman-Schwarz model
\cite{frsw}.  The rest of the terms in the transformation rules
similarly all map over appropriately.

    It is of interest to see what happens to our
Kaluza-Klein reduction Ansatz if this limit is taken.  For the metric 
Ansatz (\ref{metans}), we find that as $b$ becomes very large, the
metric becomes
%%%%%
\be
d\hat s_{11}^2 = (\ft12 b X^2)^{\fft23}\, \Big( ds_4^2 +
\fft{2}{g\, \td g}\, d\td\xi^2 + \ft12 g^{-2}\, X^{-2}\, \sum_i (h^i)^2
+ \ft12 \td g^{-2}\, X^{-2}\, \sum_i (\td h^i)^2\Big)\,,\label{d11met0}
\ee
%%%%%
where we have defined a new coordinate $\td\xi$ by
$\xi=b^{-\fft12}\, \td\xi +\ft14\pi$.  Similarly, we find that
the Ansatz for the 4-form field strength, given in 
(\ref{fans1}), (\ref{fans2}) and (\ref{fpp}), reduces to
%%%%%
\bea
\hat F_\4 &=& \fft{b}{\sqrt{2 g\, \td g}}\, \Big( X^4\, {*d\chi}\wedge
d\td\xi - \ft12 g^{-2}\, d\td\xi\wedge \ep_\3 - \ft12 \td g^{-2}\,
d\td\xi\wedge \td\ep_\3 \nn\\
&&\qquad\quad + \ft12 g^{-1}\, d\td\xi\wedge F_\2^i\wedge h^i + 
 \ft12 \td g^{-1}\, d\td\xi\wedge \wtd F_\2^i\wedge \td h^i\Big)\,.
\label{d11fans0}
\eea
%%%%%%
We see that the metric Ansatz has an overall $b^{2/3}$ constant
factor, while the 4-form Ansatz has an overall $b$ factor.  These in
fact precisely cancel out when the Ans\"atze are substituted into the
eleven-dimensional equations of motion (\ref{d11eom}), since there is
a scaling symmetry in $D=11$ under
%%%%%
\be
\hat g_{MN}\longrightarrow e^{2k}\, \hat g_{MN}\,,\qquad
\hat A_{MNP} \longrightarrow e^{3k}\, \hat A_{MNP}\,.\label{trombone}
\ee
%%%%%
Thus even though $b$ is being sent to infinity, the Ansatz still gives
a sensible limit.  In fact using (\ref{trombone}), we can effectively
set $b$ to any desired value  in (\ref{d11met0}) and (\ref{d11fans0}).  
It is convenient to take $b=2$.

   The Ans\"atze (\ref{d11met0}) and (\ref{d11fans0}) can be
reinterpreted as a first reduction step from $D=11$ to $D=10$ on the
$\td\xi$ Killing direction, followed by a reduction on $S^3\times
S^3$.  To go from $D=11$ to $D=10$ we follow the standard Kaluza-Klein
prescription, with
%%%%%
\bea
d\hat s_{11}^2 &=& e^{-\fft16\varphi}\, ds_{10}^2 +
e^{\fft43\varphi}\, (d\td\xi + \cA_\1)^2\,,\nn\\
\hat F_\4 &=& F_\4 + F_\3\wedge (d\td\xi + \cA_\1)\,,
\eea
%%%%%
where $F_\4=dA_\3 - dA_\2\wedge \cA_\1$ and $F_\3 = dA_\2$
(The field-strength reduction follows from $\hat A_\3 = A_\3 +
A_\2\wedge d\td\xi$.)  Comparing with (\ref{d11met0}) and
(\ref{d11fans0}), we see that in the $D=10$ type IIA theory we shall
have
%%%%%
\bea
ds_{10}^2 &=&(\ft{2}{g\,\td g})^{\fft18}\, \Big[
e^{\fft34\phi}\, ds_4^2 + e^{-\fft14\phi}\, \Big( g^{-2}\, \sum_i
(h^i)^2 + \td g^{-2}\, \sum_i (\td h^i)^2 \Big)\, \Big]\,,\nn\\
F_3 &=& \fft1{\sqrt{2 g\, \td g}}\, \Big[ 2 e^{2\phi}\, {*d\chi}
+ g^{-2}\, \ep_\3 + \td g^{-2}\, \td\ep_\3 -
g^{-1}\, F_\2^i\wedge h^i - \td g^{-1}\, \wtd F_\2^i\wedge
\td h^i\Big]\,,\nn\\
\varphi &=& \ft12 \phi - \ft34\log(\ft12 g\, \td g)\,,\label{d10ans}\\
F_\4 &=& 0\,,\qquad \cA_\1=0\,.\nn
\eea
%%%%%
Thus only the NS-NS fields of the type IIA theory are active (the
metric, the dilaton $\varphi$, and the 3-form field strength $F_\3$),
while the R-R fields $F_\4$ and $\cF_\2=d\cA_\1$ are zero.  The
reduction Ansatz (\ref{d10ans}) can therefore 
also be interpreted as a reduction in
the type I or heterotic string.  It is easy to check that it agrees
precisely with the reduction given in \cite{vc,cham}, for obtaining
the Freedman-Schwarz model as an $S^3\times S^3$ reduction of the
heterotic theory.  This singular limit of
the $S^7$ reduction is reminiscent of examples discussed previously in
\cite{cvlilupo}.  

    Once again, the ``one-way'' nature of the limiting procedure can
be seen in the Kaluza-Klein reduction.  One can take a singular limit
in which $S^7$ becomes $\R\times S^3\times S^3$, but one cannot
reverse the process and obtain $S^7$ as a limit of $\R\times S^3\times
S^3$.

    It is interesting to note that no analogue of the scaling
(\ref{scaling}) arises in the Freedman-Schwarz model.  As we showed,
in the standard $N=4$ $SO(4)$ gauged theory this scaling means that
there is really no distinction between the situation where the gauge
coupling constants $g$ and $\td g$ of the two $SU(2)$ factors in the
gauge group are equal or unequal.  On the other hand, the absence of
such a scaling transformation in the Freedman-Schwarz case means that
the ratio between its $SU(2)$ coupling constants $g$ and $\td g$ is a
genuine parameter of the theory, with no redefinition that maps one
value into another.

\section{Conclusion}

     In this paper, we have constructed the complete, non-linear,
explicit Kaluza-Klein Ansatz for obtaining the bosonic sector of
four-dimensional $N=4$ $SO(4)$ gauged supergravity by dimensional
reduction of eleven-dimensional supergravity on $S^7$.  Although in
principle subsumed by the $N=8$ reduction constructed in
\cite{deWitnicolai}, the advantage of our $N=4$ truncation is that the
resulting four-dimensional theory is much simpler than the maximal
$N=8$ theory, and consequently the reduction Ansatz is much more
manageable.  In fact it is because of this simplification that we have
been able to construct a fully explicit Kaluza-Klein reduction.

   The key point, in fact, is that the surviving gauge group in the
truncation, namely $SO(4)$, is the product of two $SU(2)$ factors, and
the gauge bosons for these factors arise from two separate 3-spheres
in the parameterisation of the internal 7-sphere as a foliation of
$S^3\times S^3$ hypersurfaces.  Thus we are able to benefit from the fact
that the 3-spheres are themselves $SU(2)$ group manifolds.  A group
manifold $G$ has a $G\times G$ isometry group, comprising independent
left-translations and right-translations under $G$.  Since we
need only include the gauge bosons associated with the left-translations
on each 3-sphere, the corresponding deformations of the 3-spheres
preserve their homogeneity.  Thus although the 7-sphere itself is distorted
inhomogeneously when the lower-dimensional fields are excited, these
inhomogeneities are limited to co-dimension 1, corresponding to a
distortion of the foliation whilst keeping the $S^3\times S^3$
surfaces themselves homogeneous.   For this reason, the dependence of 
the Ansatz
on the coordinates of the internal 7-sphere is restricted to the
``latitude'' coordinate $\xi$ that parameterises the foliating
surfaces.  It is still, of course,  highly non-trivial that the overall $\xi$ 
coordinate dependence cancels out in the eleven-dimensional equations
of motion under the Kaluza-Klein reduction.

    Using the results obtained in this paper, any bosonic solution of
four-dimensional $N=4$ $SO(4)$ gauged supergravity can be oxidised
back to an exact solution of eleven-dimensional supergravity.

   We have shown that the $N=4$ $SO(4)$ gauged supergravity
really has only one genuine gauge-coupling parameter, and that
although one can introduce ``independent'' parameters $g$ and $\td g$
for the two $SU(2)$ gauge groups, this is nothing but a redefinition
of fields in the theory with $g=\td g$.  This is a different situation
from the Freedman-Schwarz model, where the two gauge couplings are
genuinely independent parameters, which cannot be set equal by field
redefinitions.  

   We have also shown that the Freedman-Schwarz model arises as a
singular limit of the standard $N=4$ $SO(4)$ gauged supergravity, in
which the axion is shifted by an infinite constant, together with
appropriate rescalings of other fields.  We have shown how this
translates, in our Kaluza-Klein reduction, to a limiting case where
the 7-sphere degenerates into the product $\R\times S^3\times S^3$.

\section*{Appendices}

\appendix

\section{Derivation of the Reduction Ansatz}

    In this Appendix, we present some of the details of how we arrived
at the Kaluza-Klein metric reduction Ansatz that we present in section
2.  First, we show how the Ansatz can be deduced, in the absence of
the axion, from previous results \cite{ten} for the $U(1)^4$
truncation of four-dimensional maximal gauged supergravity.  Then, we
show how the general results in \cite{deWitnicolai} allow us to obtain
the metric Ansatz after the inclusion of the axionic scalar field.
The final part of Appendix A comprises a collection of useful {\it
lemmata} for the $SU(2)$-valued forms that are used in the
construction of the Ansatz.

\subsection{Ansatz with axion set to zero}

     The structure of the embedding of four-dimensional $N=4$ gauged $SO(4)$
supergravity in $D=11$ can be seen by first considering the maximal
abelian $U(1)^4$ embedding obtained in \cite{ten}.   In that case
there are three dilatons and three axions in the full $U(1)^4$ theory,
although in the reduction Ansatz derived in \cite{ten}, the axions
were set to zero.  We can make a further truncation to $U(1)^2$, by
setting pairs of the original four $U(1)$ gauge fields equal.  At the
same time, for consistency, two dilatons and two axions are set to
zero.  In the axion-free situation described in \cite{ten}, the metric
reduction Ansatz is
%%%%%
\be
ds_{11}^2 = \wtd\Delta^{2/3}\, ds_4^2 +g^{-2}\,
\wtd\Delta^{-1/3}\, \sum_i X_i^{-1}\, \Big( d\mu_i^2 + \mu_i^2\,
 (d\phi_i + g\,
A^i_\1)^2 \Big)\ .\label{s7metred}
\ee
%%%%%
where $\wtd \Delta = \sum_{i=1}^4 X_i\, \mu_i^2$.  The four quantities
$\mu_i$ satisfy $\sum_i \mu_i^2 =1$.  The four scalars $X_i$, which
satisfy $X_1 X_2 X_3 X_4=1$, are parameterised by the three dilatons
$\vec\phi$, as $X_i= \exp(-\ft12 \vec a_i\cdot\vec\phi)$, for certain
constant 3-vectors $\vec a_i$.  Setting two of the dilatons to zero
leads to $X_1=X_2\equiv X$, $X_3=X_4=1/X$.  At the same time, the
$U(1)$ gauge fields are set pairwise equal, with $A_\1^1=A_\1^2 \equiv
A_\1$, $A_\1^3=A_\1^4\equiv \wtd A_\1$.   Thus the metric Ansatz
(\ref{s7metred}) reduces to
%%%%%
\bea
d\hat s_{11}^2 &=& \Delta^{\fft23}\, ds_4^2 + 4 g^{-2}\,
\Delta^{\fft23}\, d\xi^2 \nn\\
&&+ g^{-2}\,\Delta^{-\fft13}\, X^{-1}\, c^2\,
\Big(d\theta^2  + \sin^2\theta\, d\varphi^2 + (d\psi+ \cos\theta\,
d\varphi - g\, A_\1)^2\Big)\nn\\
&&+ g^{-2}\,\Delta^{-\fft13}\, X\, s^2\,
\Big(d\td\theta^2  + \sin^2\td\theta\, d\td\varphi^2 
+ (d\wtd\psi+ \cos\td\theta\, d\td\varphi - g\, \wtd A_\1)^2\Big)
\,,\label{metred2}
\eea
%%%%%
where we have found it convenient to parameterise the four quantities
$\mu_i$ as
%%%%%
\be
\mu_1 = c\, \cos\ft12\theta\,,\quad \mu_2 = c\, \sin\ft12\theta\,,
\quad \mu_3 = s\, \cos\ft12\td\theta\,,\quad \mu_4 = s\, \sin\ft12\td\theta
\,,
\ee
%%%%%
where $c\equiv \cos\xi$ and $s\equiv \sin\xi$,
and the four azimuthal angles $\phi_i$ as
%%%%%
\be
\phi_1 =\ft12(\psi+\varphi)\,,\quad
\phi_2 =\ft12(\psi-\varphi)\,,\quad
\phi_3 =\ft12(\wtd\psi+\td\varphi)\,,\quad
\phi_4 =\ft12(\wtd\psi-\td\varphi)\,.
\ee
%%%%%

   If we temporarily set $X=1$ and $A_\1=\wtd A_\1=0$ (\ie turning off
the four-dimensional field excitations) , we see that the
internal seven-dimensional metric in (\ref{metred2}) becomes the round
7-sphere, written as 
%%%%%
\be
d\Omega_7^2 = d\xi^2 + \cos^2\xi\, d\Omega_3^2 + \sin^2\xi\,
d\wtd\Omega_3^2\,,\label{7133met}
\ee
%%%%%
where $d\Omega_3^2$ and $d\wtd\Omega_3^2$ are two separate unit
3-sphere metrics, written in terms of the Euler angles $(\theta,
\varphi, \psi)$ and $(\td\theta,\td\varphi, \wtd\psi)$ respectively.

    A natural generalisation of the reduction Ansatz (\ref{metred2})
now suggests itself, in which we enlarge the $U(1)$ gauge field in
each 3-sphere to $SU(2)$:
\be
d\hat s_{11}^2 = \Delta^{\fft23}\, ds_4^2 + 4 g^{-2}\,
\Delta^{\fft23}\, d\xi^2 + g^{-2}\,\Delta^{-\fft13}\,\Big( X^{-1}\, c^2\,
 \sum_{i=1}^3 (h^i)^2 + X\, s^2\,   \sum_{i=1}^3 (\td h^i)^2\Big)\,, 
\label{metred3}
\ee
%%%%%
where 
%%%%%
\be
h^i\equiv \sigma_i -g A_\1^i\,,\qquad  
\td h^i\equiv \td\sigma_i -g \wtd A_\1^i\,,
\ee
%%%%%
where $\sigma_i$ are left-invariant 1-forms on the first $S^3$, and
$\wtd\sigma_i$ are left-invariant 1-forms on the second $S^3$. 

\subsection{Ansatz with non-vanishing axion}

    In the above, we considered the situation when the axion of the
$N=4$ theory is set to zero.  When the axion is non-zero, we cannot
deduce the form of the metric Ansatz from the previous results in
\cite{ten}.  Now, we can make use of the general formalism in
\cite{deWitnicolai}, where the reduction Ansatz for the $N=8$ theory
was obtained.  In particular, the full metric Ansatz in
\cite{deWitnicolai} is relatively simple, and by truncating it 
appropriately we are able to construct the Ansatz for the $N=4$ theory.

   To begin, we need to determine the tensors $u_{ij}{}^{IJ}(x)$ and
$v_{ij IJ}(x)$ that appear in the definition of the scalar 
56-bein $\cv$ and its inverse,
%%%%%
\be
\cv =\pmatrix{ u_{ij}{}^{IJ} & v_{ijKL}\cr
               v^{k\ell IJ} & u^{k\ell}{}_{KL}}\,,\qquad
\cv^{-1} =\pmatrix{ u^{ij}{}_{IJ} & -v_{k\ell IJ}\cr
               -v^{ij KL} & u_{k\ell}{}^{KL}}\,.
\ee
%%%%%
In the $N=4$ gauged $SU(2)\times SU(2)$ truncation of the full $N=8$
gauged $SO(8)$ theory, we find that these are given by
%%%%%
\bea
&&u_{ab}{}^{cd} = 2 \cosh\ft12\lambda\, \delta_{ab}^{cd}\,, \qquad
u_{\bar a\bar b}{}^{\bar c\bar d} = 
2 \cosh\ft12\lambda\, \delta_{\bar a\bar b}^{\bar c\bar d}\,,\qquad
u_{a\bar b}{}^{c\bar d} = 2 \delta_a^c\, \delta_{\bar b}^{\bar
d}\,,\nn\\
&&v_{abcd} = \sinh\ft12\lambda\, e^{\im \sigma}\, \ep_{abcd}\,,\qquad
v_{\bar a\bar b\bar c\bar d} =
\sinh\ft12\lambda\, e^{-\im \sigma}\, 
\ep_{\bar a\bar b\bar c\bar d}\,,\label{uvexp}
\eea
%%%%%
where we have split the indices $i=(1,8)$ into $i=(a,\bar a)$, where $a=(1,4)$
and $\bar a=(5,8)$, and similarly for $I$.  The fields $\lambda$ and
$\sigma$ are related to the usual dilaton $\phi$ and axion $\chi$ by
%%%%%
\bea
\cosh\lambda &=& \cosh\phi + \ft12 \chi^2\, e^\phi\,,\nn\\
\cos\sigma \,\sinh\lambda &=& \sinh\phi - \ft12 \chi^2\,
e^\phi\,,\label{siglam}\\
\sin\sigma \,\sinh\lambda &=& \chi\, e^\phi\,.\nn
\eea
%%%%%
(This is the mapping from the metric $ds_2^2 = d\lambda^2 +
\sinh^2\lambda\, d\sigma^2$ to $ds_2^2 = d\phi^2 + e^{2\phi}\,
d\chi^2$.  Note that in terms of $\sigma$ and $\lambda$, the scalar
potential (\ref{pot2}) is simply given by $V=-4 g^2\, (\cosh\lambda + 2)$.)  

    It is shown in \cite{wna,deWitnicolai} that the Ansatz 
for the inverse metric in the internal space (the 7-sphere) is
%%%%%
\be
\hat\Delta(x,y)\, g^{mn}(x,y) = \ft12 
(K^{m IJ}\, K^{n KL} + K^{n IJ}\, K^{m KL})\, (u_{ij}{}^{IJ} + v_{ij
IJ})\, (u^{ij}{}_{KL} + v^{ijKL})\,,\label{dwnmet}
\ee
%%%%%
where
%%%%%
\be
\hat\Delta^2 = \fft{\det(g_{mn}(x,y))}{\det(\bar g_{mn}(y))}\,.
\ee
%%%%%
Here $\bar g_{mn}(y)$ denotes the metric of the undistorted round
7-sphere, and $K^{m IJ}$ are the 28 Killing vectors in this metric.  
Substituting our expressions (\ref{uvexp}) into (\ref{dwnmet}), we
find
%%%%%
\bea
\hat\Delta\, \bar g^{mn}(x,y)  &=& \sum_{i,j} K^{m\, ij}\, K^{n\, ij} 
+ \ft14 (X^2-1)\,
\sum_{\a=1}^3\Big( (J^\a_{ab}\, K^{m\, ab})^2 + (J^\a_{\bar a\bar b} \,
K^{m\, \bar a\bar b})^2 \Big)\nn\\
&& + \ft14 (\tX^2-1)\,
\sum_{\a=1}^3\Big( (\wtd J^\a_{ab}\, K^{m\, ab})^2 
+ (\wtd J^\a_{\bar a\bar b} \,
K^{m\, \bar a\bar b})^2 \Big)\,,\label{inversemet}
\eea
%%%%%
where
%%%%%
\bea
&&J^1_{12}=J^1_{34}=J^2_{13}=-J^2_{24}=J^3_{14}=J^3_{23}=1\,,\nn\\
&&J^1_{56}=J^1_{78}=J^2_{57}=-J^2_{68}=J^3_{58}=J^3_{67}=1\,,\nn\\
&&\wtd J^1_{12}=-\wtd J^1_{34}=\wtd J^2_{13}=\wtd J^2_{24}
=\wtd J^3_{14}=-\wtd J^3_{23}=1\,,\\
&&\wtd J^1_{56}=-\wtd J^1_{78}=\wtd J^2_{57}=\wtd J^2_{68}
=\wtd J^3_{58}=-\wtd J^3_{67}=1\,.\nn
\eea
%%%%%
Thus $J^\a_{ab}$ and $J^\a_{\bar a\bar b}$ are self-dual in the $4+4$
dimensional subspaces spanned by $i=(a,\bar a)$, while   
$\wtd J^\a_{ab}$ and $\wtd J^\a_{\bar a\bar b}$ are anti-self-dual.  

   It is easy to see that the 3 Killing vector combinations $K^{m\, \a}
\equiv J^\a_{ab} \, K^{m\,ab}$ and the 3 combinations $\bar K^{m\,
\a}\equiv J^\a_{\bar a\bar b} \, K^{m\,\bar a\bar b}$ each close on
$SU(2)$, and that the two sets mutually commute.  Likewise, the 3
combinations $\wtd K^{m\, \a} \equiv \wtd J^\a_{ab} \, K^{m\,ab}$ and
the 3 combinations $\wtd{\bar K}^{m\, \a}\equiv \wtd J^\a_{\bar a\bar b}
\, K^{m\,\bar a\bar b}$ each close on $SU(2)$, and these commute with
each other and the other two $SU(2)$'s.  In fact what we are seeing
here are the sets of $3+3$ Killing vectors on each of two 3-spheres:
$K^{m\, \a}$ and $\bar K^{m\,\a}$ are the left-translation and
right-translation Killing vectors of one 3-sphere, while $\wtd K^{m\,
\a}$ and $\wtd{\bar K}^{m\, \a}$ are the left-translation and
right-translation Killing vectors of the other 3-sphere.  

   Now, considering the first $S^3$, the sum of the
squares of the left-translation Killing vectors, $K^{m\, \a}\, K^{n\,
\a}$ is equal to the sum of the squares of the right-translation
Killing vectors, $\bar K^{m\,\a}\,\bar K^{n\,\a}$, each sum 
giving the bi-invariant inverse metric $g_3^{mn}$ on the $S^3$.
A similar remark applies to the second $S^3$. 
Also, the sum of the squares of all 28 Killing vectors gives the
inverse metric on the round 7-sphere, so (\ref{inversemet}) becomes
%%%%%
\be
\hat\Delta\, \bar g^{mn}(x,y)  = \bar g^{mn}(y)
+  (X^2-1)\, g_3^{mn}(y) +
 (\tX^2-1)\,  \td g_3^{mn}(y)\,.\label{inversemet2}
\ee
%%%%%
This metric is easily inverted, and in terms of the representation
$d\Omega_7^2$ in (\ref{7133met}) for the round 7-sphere metric, and
$d\Omega_3^2 = \ft14 \sum_{i=1}^3 \sigma_i^2$ and 
$d\wtd \Omega_3^2 = \ft14 \sum_{i=1}^3 \td\sigma_i^2$ for the two
round 3-sphere metrics, we obtain
%%%%%
\be
ds_7^2 = \hat \Delta^{-1}\, \Big( d\xi^2 + \ft14 \fft{c^2}{c^2 \, X^2 +
s^2}\,\sum_{i=1}^3 \sigma_i^2 + \ft14 \fft{s^2}{s^2 \, \tX^2 +
c^2}\,\sum_{i=1}^3 \td\sigma_i^2\Big)\,,
\ee
%%%%%
where, as usual, $c=\cos\xi$ and $s=\sin\xi$.
From the expression in \cite{wna,deWitnicolai} for the 
eleven-dimensional metric in
terms of the seven-dimensional one, $d\hat s_{11}^2 =
\wtd\Delta^{-1}\, ds_4^2 + ds_7^2$, and noting that our factor
$\Delta$ is related to the corresponding factor $\hat\Delta$ of
\cite{wna,deWitnicolai} by $\Delta = \hat\Delta^{-3/2}$, 
we eventually arrive at our
metric Ansatz (\ref{metans}), after introducing the $SO(4)=
SU(2)\times SU(2)$ gauge fields as described earlier.  
Note that it reduces to (\ref{metred3})
if we set the axion to zero.  (We have introduced the gauge coupling
constant by means of appropriate rescalings.)

   In order to check the consistency of the reduction Ans\"atze
presented in section 2, a necessary ingredient is the calculation of
the Ricci tensor for the metric Ansatz (\ref{metans}).  If we define
%%%%%
\be
e^{\beta}\equiv \Delta^{\fft13}\,,\qquad
e^\gamma \equiv (\sqrt2\, g)^{-1}\, c\, 
\Delta^{\ft13} \,\Omega^{-\fft12}\,,\qquad
e^{\td\gamma} \equiv (\sqrt2\, g)^{-1}\, s\, 
\Delta^{\ft13} \,\wtd\Omega^{-\fft12}\,,
\ee
%%%%%
then a natural orthonormal basis is
%%%%%
\be
\hat e^a = e^\beta\, e^a\,,\qquad \hat e^0 = \sqrt2\,
g^{-1}\,e^\beta\, d\xi\,,\quad
\hat e^i = e^\gamma\, h^i\,,\quad
\hat e^{\td i} = e^{\td \gamma}\, \td h^i\,.
\ee
%%%%%
In terms of this basis, we find that the vielbein components of the
Ricci tensor are given by
%%%%%
\bea
\hat R_{00} &=& \Delta^{-\fft23}\, \Big[ -\square \beta + \beta_a\, \beta_a + 
\ft12 g^2( -4\beta''-3 \gamma'' -3\td\gamma'' + 3 \beta'\, \gamma' + 3 \beta'\,
\td \gamma' - 3{\gamma'}^2 - 3({\td\gamma'})^2)\Big]\,,\nn\\
\hat R_{0a} &=& 
\ft3{\sqrt2} g\, \Delta^{-\fft23}\, \Big[ (\beta_a-\gamma_a)\,
\gamma' + (\beta_a-\td\gamma_a)\, \td\gamma'\Big]\,,\nn\\
\hat R_{0i} &=& 0\,,\qquad \hat R_{0\td i} = 0\,,\nn\\
\hat R_{ab} &=& \Delta^{-\fft23}\, \Big[ R_{ab} - 3(\beta_a\, \beta_b +
\gamma_a\, \gamma_b + \td\gamma_a\, \td\gamma_b) +
(-\square \beta + 2\beta_c\, \beta_c)\, \eta_{ab} \nn\\
&&-\ft14 c^2\, \Omega^{-1}\,  
F_{ac}^i\, F_{bc}^i -\ft14 s^2\, \wtd\Omega^{-1}\, 
\wtd F_{ac}^i\, \wtd F_{bc}^i 
-\ft12 g^2 (\beta'' + 6 \beta'\, \cot 2\xi)\, \eta_{ab} \Big]\,,\nn\\
\hat R_{ai} &=& -\ft1{2\sqrt2}\, 
c\, \Delta^{-\fft23}\, \Omega^{-\fft12}\, [
D_b\, F^i_{ab} -2 (\beta_b-\gamma_b)\, F^i_{ab}]\,,\nn\\
\hat R_{a\td i} &=& -\ft1{2\sqrt2}\, 
s\, \Delta^{-\fft23}\, \wtd\Omega^{-\fft12}\, [
\wtd D_b\, \wtd F^i_{ab} -2 (\beta_b-\td\gamma_b)\, \wtd F^i_{ab}]\,,\nn\\
\hat R_{ij} & =& \Delta^{-\fft23}\, \Big[ \ft12 g^2\, (-\gamma'' - 6
\gamma'\, \cot 2\xi + 2\Omega\, c^{-2})\, \delta_{ij} -\square
\gamma\, \delta_{ij} +
\ft1{8} c^2\, \Omega^{-1}\, F^i_{ab}\, F^j_{ab}\Big]\,,\nn\\
\hat R_{\td i\td j} & =& \Delta^{-\fft23}\, \Big[ \ft12 g^2\, 
(-\td\gamma'' - 6
\td\gamma'\, \cot 2\xi + 2\wtd\Omega\, s^{-2})\, \delta_{ij} -\square
\td\gamma\, \delta_{ij}+ 
\ft1{8} s^2\, \wtd\Omega^{-1}\, \wtd F^i_{ab}\, \wtd F^j_{ab}\Big]\,,\nn\\
\hat R_{i\td j} &=& \ft18 s\, c\, \Delta^{-\fft53}\, 
F^i_{ab}\, \wtd F^j_{ab}\,.\label{ricci}
\eea
%%%%%
Note that here $\beta_a$, $\gamma_a$ and $\td\gamma_a$ denote the vielbein
components of the four-dimensional spacetime derivatives of these
functions, so $\beta_a=\del_a\, \beta$, {\it etc}.  Similarly, 
$\beta'$, $\gamma'$
and $\td\gamma'$ denote their derivatives with respect to $\xi$.
Useful identities are $\beta_a+\gamma_a+\td\gamma_a=0$, and $\beta'+\gamma' +
\td\gamma' = 2\cot 2\xi$. 

    Some partial formulae for the 4-form Ansatz are presented in
\cite{deWitnicolai}, but it is difficult to turn them into explicit
expressions, and in any case not all components are presented.  We
therefore determined the 4-form Ansatz in section 2 by brute-force methods.

\subsection{Some $SU(2)$ Lemmata}

   Here, we collect together some useful properties of
the $SU(2)$-valued forms that arise in the reduction Ansatz.  We define
$h^i\equiv \sigma_i-g\, A_\1^i$, and $\td h^i \equiv \td\sigma_i - g\,
\wtd A_\1^i$, where $\sigma_i$ and $\td\sigma_i$ are sets of
left-invariant 1-forms on the two 3-spheres, satisfying
(\ref{leftinv}).  The Yang-Mills field strengths are defined by
(\ref{fde0}). From their Bianchi identities, we see that we should
define gauge-covariant exterior derivatives
%%%%%
\be
D \omega^i \equiv d\omega^i + g\, \ep_{ijk}\, A_\1^j\wedge
\omega^k\,,\qquad
\wtd D \td\omega^i \equiv d\td\omega^i + g\, \ep_{ijk}\, \wtd A_\1^j\wedge
\td\omega^k\,,\label{gaugecov}
\ee
%%%%%
for any forms with an adjoint index of the untilded or tilded
$SU(2)$.  The Bianchi identities themselves are then $D\, F_\2^i=0$,
$\wtd D\, \wtd F_\2^i=0$.  

    We can now derive a number of lemmata.  Since every formula for the
untilded $SU(2)$ has an identical companion formula for the tilded
$SU(2)$, we shall just present the untilded ones.
%%%%%
\be
dh^i = -\ft12\ep_{ijk}\, h^j\wedge k^k - g\, F_\2^i - g\, \ep_{ijk}\,
A_\1^j\wedge h^k\,.\label{eq1}
\ee
%%%%%
This can be written more elegantly using the gauge-covariant exterior
derivative defined in (\ref{gaugecov}):
%%%%%
\be
D h^i =  -\ft12\ep_{ijk}\, h^j\wedge h^k - g\, F_\2^i\,.\label{eq2}
\ee
%%%%%
This is a convenient way to express the result, because the
gauge-covariant exterior derivative respects the Leibniz rule, just
as the ordinary exterior derivative does.  Thus, for example,
%%%%%
\bea
d(h^i\wedge F_\2^i) &=& Dh^i\wedge F_\2^i - h^i\wedge DF_\2^i
=  Dh^i\wedge F_\2^i \nn\\
&=& -\ft12\ep_{ijk}\, h^j\wedge h^k\wedge F_\2^i - 
g\, F_\2^i\wedge F_\2^i\,.
\eea
%%%%%
Another result is
%%%%%
\be
d(h^i\wedge {*F_\2^i}) =  -\ft12\ep_{ijk}\, h^j\wedge h^k\wedge {*F_\2^i} - 
g\, {*}F_\2^i\wedge F_\2^i - h^i\wedge D{*F_\2^i}\,.
\ee
%%%%%

   It is also useful to derive that
%%%%%
\be
D(\ep_{ijk}\, h_j\wedge h_k) = -2\ep_{ijk}\, h^j\wedge Dh^k = 2g\,
\ep_{ijk}\,h^j\wedge F_\2^k\,.\label{eq3}
\ee
%%%%%
From this, we see, for example, that
%%%%%
\be
d(\ep_{ijk}\, h_i\wedge h_j\wedge F_\2^k) = 0\,,\qquad
d(\ep_{ijk}\, h_i\wedge h_j\wedge {*F_\2^k})= 
\ep_{ijk}\, h_i\wedge h_j\wedge D{*F_\2^k}\,.\label{eq4}
\ee
%%%%%
We also have the result that with $\ep_\3\equiv h^1\wedge h^2\wedge
h^3$, 
%%%%%
\be
d\ep_\3 = -\ft12 g\, \ep_{ijk}\, h^i\wedge h^j\wedge F_\2^k\,.\label{eq5}
\ee
%%%%%

\section*{Acknowledgement}

    We are grateful to Mikhail Shifman and Lenny Susskind for useful
discussions about consistent truncations, and to Jim Liu for
discussions about gauged supergravities.  M.C. is grateful to the
Physics Department at Texas A\&M University for hospitality during
part of this work.

\end{document}